# 環境適応型ソフトウェア実現に向けた検討

山登庸次†　　野口博史†　　片岡操†　　磯田卓万†

† NTT ネットワークサービスシステム研究所，東京都武蔵野市緑町 3-9-11
E-mail: †yamato.yoji@lab.ntt.co.jp

**あらまし** 近年，GPU や FPGA 等のヘテロなハードウェアの利用が増えており，IoT デバイスも増えている．しかし，ヘテロなハードウェアの活用には，技術的ハードルが高い．そこで，将来的には，開発者は行いたいロジックだけ記述すると，ソフトウェアが環境に適応し，ヘテロなハードウェアを活用できることが必要になると考える．そこで，本稿では，アプリケーションを一度書けば，配置される場所の環境に応じて，GPU や FPGA，IoT デバイスを扱えるようにコードが変換，自動設定され，高い性能で運用できる，環境適応型ソフトウェアを提案する．環境適応型ソフトウェアの実現に向けた処理フローと要素技術を説明する．
**キーワード** 環境適応型ソフトウェア, GPGPU，FPGA，自動オフロード，性能

## Study to achieve environment adaptive software

Yoji YAMATO†, Hirofumi NOGUCHI†, Misao KATAOKA†, and Takuma ISODA†

† Network Service Systems Laboratories, NTT Corporation, 3-9-11, Midori-cho, Musashino-shi, Tokyo
E-mail: †yamato.yoji@lab.ntt.co.jp

**Abstract** Recently, heterogeneous hardware such as GPU and FPGA is used in many systems and also IoT devices are increased repidly. However, to utilize heterogeneous hardware, the hurdles are high because of much technical skills. In order to break down such a situation, we think it is required in the future that application programmers only need to write logics to be processed, then software will adapt to the environments with heterogeneous hardware, to make it easy to utilize heterogeneous hardware and IoT devices. Therefore, in this paper, we propose environment adaptive software to operate an once written application with high performance by automatically converting the code and configuring setting so that we can utilize GPU, FPGA and IoT devices in the location to be deployed. We explain a proceesing flow and elemental technologies to achieve environment adaptive software.
**Key words** Environment Adaptive Software, GPGPU, FPGA, Automatic Offloading, Performance.

## 1. はじめに

近年，ムーアの法則が終焉し，CPU が 1.5 年で 2 倍の密度になることが期待できなくなるのではと言われている．その状況も踏まえて，CPU だけでなく，GPU（Graphics Processing Unit）や FPGA（Field Programmable Gate Array）といったヘテロジニアスなハードウェアをシステムに用いることが増えている．例えば，Amazon は，GPU，FPGA のインスタンス [1] をクラウド技術（例えば，[2] [3] [4]）を使って提供しており，Microsoft は FPGA を使って Data Center の効率を高めている [5]．

しかし，ヘテロなハードウェアをアプリケーションで適切に活用するためには，ハードウェアに合わせたプログラミングや設定が必要で，CUDA（Compute Unified Device Architecture）[6]，OpenCL（Open Computing Language）[7] といった技術知識が求められ，多くのプログラマーにとってハードルは高い．

IoT（Internet of Things）技術（例えば，[8]- [13]）の進展に伴い，IoT デバイスは増加を続けており，2020 年には数百億，2020 年代後半には兆に達するデバイスがネットワークに繋がると言われている．IoT の応用分野は，製造，流通，医療，農業等多岐に渡り，サービス連携技術等 [14]- [21] を用いて，製品の製造過程を可視化するといった，多彩なアプリケーションが出ている．

IoT アプリケーションで，IoT デバイスのより細かい制御を行う際は，組込みソフトウェアの知識やアセンブリ等が必要になる．ゲートウェイ（GW）で，複数の IoT デバイスを集約して制御することも数多くされているが，ゲートウェイのリソースは限られるため，利用する環境に応じた設計が必要である．

整理すると，GPU や FPGA 等のヘテロなハードウェア，多くの IoT デバイスを活用したアプリケーションの期待は高まっ

ているが，それらの活用にはハードルが高いのが現状である．そこで，そのような現状を打破するため，ヘテロなハードウェアや，IoT デバイスを容易に活用できるようにするため，アプリケーションプログラマーは処理したいロジックだけを書き，使う先の環境（ヘテロハードウェアや利用する IoT デバイス）に合わせて，ソフトウェアが適応して，環境にあった動作をすることが，将来的に求められると考える．

1995 年に登場した Java [22] は，一度書いたソフトウェアを，別のマシンでも動作できるようにする，環境適応のパラダイムシフトを起こした．しかし，移行した先での性能については，特に考慮がされていなかった．私達は，一度書いたソフトウェアを，配置先環境で GPU や FPGA，IoT デバイス等を活用できるように，コード変換等を自動で行う事で，高性能でアプリケーションを動作させることを目標にした，環境適応型ソフトウェアを提案する．

## 2. 既存関連技術

環境適応ソフトウェアという点では，Java が挙げられる．Java は，Java Virtual Machine という仮想実行環境を用いて，一度書いたソフトは，再度のコンパイル不要で，異なる OS のマシンでも動作可能にしている（Write Once, Run Anywhere）．しかし，移植した先で，期待する性能が出るかは考慮がされておらず，性能チューニングや，移植した先でのデバッグ等の稼働は小さくないことが課題であった（Write Once, Debug Everywhere）．

GPU の計算能力を画像処理以外にも使う GPGPU（General Purpose GPU）のための開発環境 CUDA が発展している．CUDA は GPGPU 向けの開発環境だが，GPU，FPGA，メニーコア CPU 等のヘテロハードウェアを統一的に扱うための標準規格として OpenCL も登場している．CUDA や OpenCL では，C 言語の拡張によるプログラミングを行うが，GPU 等のデバイスと CPU の間のメモリコピー，解放等を記述する必要があり，記述の難度は高い．

簡易に GPGPU を行うため，ディレクティブベースで，並列処理すべき個所を指定し，ディレクティブに従いコンパイラがデバイス向けコードに変換する技術が有る．技術仕様として，OpenACC [23] 等，コンパイラとして PGI コンパイラ [24] 等がある．OpenACC は C/C++向けであるが，IBM の Java JDK [25] は，ラムダ記述に従い GPU オフロード処理が可能である．

このように，CUDA, OpenCL, OpenACC 等の技術により，GPU や FPGA へのオフロードが可能になっている．しかし，GPU 処理は行えるようになっても，高速化には課題が多い．マルチコア CPU 向けには，例えば，Intel コンパイラ [26] 等の自動並列化機能を持つコンパイラがある．自動並列化時は，プログラム上の for 文等の並列可能部を抽出するが，GPU を用いる場合は，CPU-GPU メモリ間のデータ転送オーバヘッドのため性能が出ないことも多い．GPU を用いて高速化する際は，スキル者が，CUDA でのチューニングや，PGI コンパイラ等で適切な並列処理部を探索することが必要になっている．

このため，スキルが無いユーザが GPU や FPGA を使ってアプリケーションを高性能化することは難しいし，自動並列化技術等を使う場合も並列処理可否の試行錯誤や高速化できない場合があった．

IoT デバイスに関しては，計算リソース等が限られている IoT デバイスでは，細かい制御を行う際は，アセンブリ等の組み込みソフトウェアの知識が必要になるのが現状である．Rasberry Pi 等のシングルボードコンピュータでは，リソース量は限られるものの，Linux や Java 等が動作するため，Rasberry Pi を GW として複数の IoT デバイスからデータを収集したり制御したりする等の自由度が開発者に出てくる．しかし，IoT デバイスを何台収容するかや，IoT デバイスとシングルボードコンピュータでどのように処理を分担するか等は，アプリケーション，利用形態によって異なり，環境に合わせた設計が必要である．

## 3. 環境適応型ソフトウェアの提案

### 3.1 環境適応型ソフトウェアで考慮する要素

近年，サービス連携（[27] [28] [29] 等）で外部コンポーネントと連携してアプリケーションを作り，クラウドファーストという言葉もあるように，作ったアプリケーションをクラウド等の事業者設備で動作させることは一般的となっている．その際に，ユーザはアプリケーションを低コストで高性能に運用することを求めている．アプリケーションを動作させる場合にコストや性能に大きく影響する点として以下の考慮が必要である．

まず，GPU や FPGA 等のアクセラレータを使う方が性能，コストで効果がある場合にそれらの利用が考えられる．勿論，通常の CPU 向けに作られたコードではそれらのアクセラレータは使えないため，画像処理や FFT（Fast Fourier Transform）処理といった，GPU や FPGA に適した処理を，それらハードウェアにオフロードするコードに変換やライブラリ呼び出しを行う必要がある．コード変換は，IoT デバイスの制御等の処理を，Rasberry Pi のようなシングルボードコンピュータに切り出して，配置する際も必要である．

動作させるアプリケーションのコードが決まると，どの程度のリソース量を確保するかの決定が必要である．例えば，CPU と GPU で動作させるアプリケーションの場合に，CPU 処理が 1000 秒で，GPU 処理が 1 秒等の場合は，仮想マシンの CPU リソースを増強した方がシステムとして性能が出ることが期待できる．

アプリケーションを実行する場所も性能に影響する．例えば，IoT カメラの画像分析で不審者探索を 0.5sec 以内に行いたい場合に，クラウドまでデータを上げてから画像分析しては遅延が大きくなるので，カメラデータを集約するゲートウェイや，NW の端点であるエッジサーバで画像分析することが必要になる等，処理場所の考慮が必要である．また，画像分析をエッジサーバで行い，不審者がいる場合だけ詳細画像をクラウドに送る場合でも，処理場所によって，計算量，通信トラフィックが変わるため，コストも変化する．

ハードウェアに合わせたコード変換，リソース量調整，配置

場所調整が終わり，アプリケーションの運用を開始しても，運用中にリクエスト特性が大きく変わった場合など，開始当初の性能が保てなくなる場合がある．そういった際は，運用中に構成を変更することで，システムとしての性能，コストを改善することも考慮が必要である．

以上の通り，環境適応ソフトウェアでは，コード変換，リソース量調整，配置場所調整，運用中再構成の4つを，対象とする．

### 3.2 環境適応型ソフトウェアの処理フロー

前サブ節で上げた点を考慮した環境適応を実現するため，図1を用いて，以下の処理フローを提案する．環境適応型ソフトウェアは，環境適応機能，テストケースDB，コードパターンDB，設備リソースDB，検証環境，商用環境からなる機能が連携して実現される．

Step1 コード分析：

ユーザは動作させたいアプリケーションコードと利用を想定したテストケース，要望する性能とコストを，環境適応機能に指定する．環境適応機能は，アプリケーションのコードを分析する．分析では，ループ文や変数の参照関係，処理する機能ブロック（FFT処理）等，コードの構造を把握する．

Step2 オフロード可能部抽出：

環境適応機能は，アプリケーションコードの並列処理可能なループ文やFFT処理の機能ブロック等，オフロード可能な処理をコードパターンDBを参照して特定し，オフロード先に応じた中間言語（OpenCL等）を抽出する．なお，中間言語抽出は一度で終わりでなく，適切なオフロード領域探索のため，実行試行して最適化するため反復がされる．

Step3 適切なオフロード部探索：

次に，環境適応機能は，検証用環境として，GPUやFPGA，IoTデバイス用GW等を備えた検証用環境に，中間言語から導かれる実行ファイルをデプロイする．配置したファイルを起動し，想定するテストケースを実行して，オフロードした際の性能を測定する．ここで，GPUやFPGA等オフロード先に応じて，この性能測定結果を用いて，より適切なオフロードとするため，Step2の中間言語抽出のステップに戻り，別パターンの抽出を行い，性能測定を試行する．検証環境での性能測定を繰り返し，最終的にデプロイするコードパターンを決定する．

Step4 リソース量調整：

Step3でコードパターンを決定後は，環境適応機能は，適切なリソース量の設定を行う．Step3の検証環境性能測定で取得される，想定するテストケースの処理時間の中で，CPU処理時間とGPU等のCPU以外ハードウェアの処理時間を分析し，適切なリソース比（CPUとGPU等ハードウェアの確保するリソースの比．例：vCPUコア4：仮想GPU1が適切等）を定める．次に，ユーザの要望する性能，コストと適切なリソース比を鑑みて，実際に確保するリソース量を定める（例：vCPUコア8，仮想GPU2が，性能，コストを満たす量）．

Step5 配置場所調整：

Step3で定めたコードパターンの実行ファイルを，Step4で定めたリソース量を確保して配置する際に，環境適応機能は，性能，コストが適切になる場所を計算し配置先を決める．実行するアプリケーションの想定するテストケースの特性，設備リソースDBの情報から，性能とコストが適切になる配置場所を計算する．例えば，IoTカメラの画像情報を分析して不審者を見つける処理を，0.5sec以内の遅延で行いたいような場合は，IoTカメラに近いエッジサーバを特定して，配置する．ここで，配置したい場所には，リソース量制限から，必要なリソースを確保できない場合等は，リソース量や場所を再調整するため，Step4に処理を戻す場合がある．

Step6 実行ファイル配置と動作検証：

Step5で定めた商用環境配置場所に，Step3で定めたコードパターンの実行ファイルを，Step4で定めたリソース量を確保して配置すると，期待通りの動作となるかを，環境適応機能は，動作検証試験を行う．ユーザが指定した想定テストケースや，テストケースDBに保持されているアプリケーションリグレッションテストケースを用いて，動作検証する．この際に，想定テストケースの商用環境での実際の性能を，確保した全リソースのスペックやコストも含めて，ユーザに提示し，ユーザにサービス開始判断をもらい，OKの場合にアプリケーションの運用を開始する．

Step7 運用中再構成：

Step6で開始したアプリケーション運用にて，リクエスト特性変化等で当初期待していた性能が出ない場合に，環境適応機能は，ソフトウェア設定，ソフトウェア/ハードウェア構成を再構成する．ソフトウェア設定とは，リソース量や配置場所の再変更を意味しており，例えば，CPUとGPUの処理時間バランスが悪い場合に，リソースの比を変更したり，リクエスト量が増え応答時間が劣化してきた場合に，リソースの比はキープして量を増やす．あるいは，配置する場所を別のクラウドに変えるなどである．ソフトウェア/ハードウェア構成とは，コード変換から行い，GPUであればオフロード処理するロジックを変更したり，FPGAのようにハードウェアロジックを運用中に変更できる場合はハードウェアロジックを再構成することを意味している．例えば，後者で，SQL DBとNo SQLのDBを両運用している場合に，元々SQLリクエストが多かったが，NoSQLリクエストが一定よりも増えてきた場合に，NoSQLをアクセラレートするFPGAにロジックを再構成する．

ここで，Step1-7で，環境適応に必要な，コード変換，リソース量調整，配置場所調整，運用中再構成を一括して行う処理フローを説明したが，行いたい処理だけ切出すことも可能である．例えば，GPU向けにコード変換だけ行いたい場合は，Step1-3だけ行い，環境適応機能や検証環境等必要な部分だけ利用すれば良い．

## 4. 環境適応ソフトウェアの要素技術

3節提案のフローを実現する要素技術を説明する．

### 4.1 コード変換

Step1のコード分析については，Clang [30] 等の構文解析ツールを用いて，アプリケーションコードの分析を行う．コード分析は，オフロードするデバイスを想定した分析が必要になるため，一般化は難しいが，ループ文や変数の参照関係等のコード

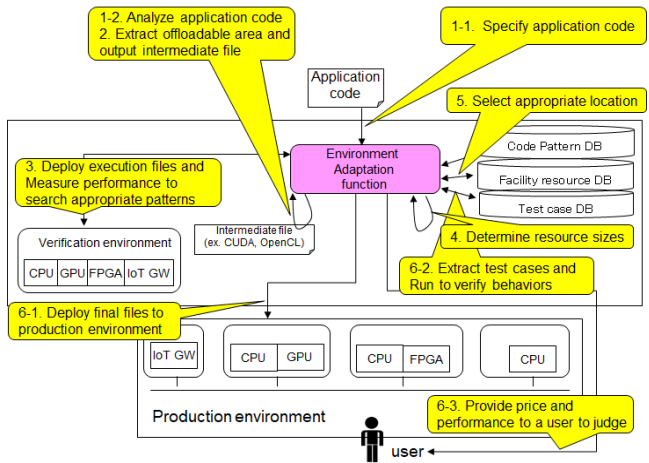

図 1　環境適応ソフトウェア処理フロー

の構造を把握したり，機能ブロックとして FFT 処理を行う機能ブロックであることや，FFT 処理を行うライブラリを呼び出している等を把握する．機能ブロックの判断は，Machine が自動判断する事は難しいが，CCFinderX [31] 等の類似コード検出ツールを用いて類似度判定等で把握する．また，Clang は C/C++向けツールであるが，解析する言語に合わせたツールを選ぶ必要がある．

Step2-3 は処理をオフロードする，GPU，FPGA，IoT デバイス制御用 GW 等，オフロード先に合わせた検討が必要となる．一般に，性能に関しては，最大性能になる設定を一回で自動発見するのは難しいため，オフロードパターンを何度か性能測定を検証環境で繰り返し試行し，高速化できるパターンを見つけることを行う．

本サブ節では既に検討が進んでいる，GPU 向けオフロードと FPGA 向けオフロードについて説明する．

**4.1.1　GPU 向けオフロード**

アプリケーションは，多種多様だが，映像分析のための画像処理（[32] 等），センサデータ分析のための機械学習等，計算量が多いアプリケーションでは，繰り返しが多い．そこで，アプリケーションのループ文を GPU に自動でオフロードする事での高速化を行う．

しかし，2 節で記載の通り，高速化には適切な並列処理が必要である．特に，GPU を使う場合は，CPU と GPU 間のメモリ転送のため，データサイズやループ回数が多くないと性能が出ないことが多い．また，メモリプロセスの持ち方やメモリデータ転送のタイミング等により，並列高速化できる個々のループ文の組み合わせが，最速とならない場合等がある（例：10 個の for 文で，1 番，5 番，10 番の 3 つが CPU に比べて高速化出来る場合に，1 番，5 番，10 番の 3 つの組み合わせが最速になるとは限らない等）．

そこで，著者の以前の研究である [33] では，CPU 向け汎用プログラムから，自動で適切なオフロード領域を抽出するため，並列可能ループ文群に対して遺伝的アルゴリズム（GA）を用いて検証環境で性能検証試行を反復し適切な領域を探索する．遺伝子の部分の形で，高速化可能な並列処理パターンを保持し組み換えていくことで，取り得る膨大な並列処理パターンから，効率的に高速化可能なパターンを探索を狙っている．GA は，生物の進化過程を模倣した組合せ最適化手法の一つで，GA のフローチャートは，初期化→評価→選択→交叉→突然変異→終了判定である [34]．

しかし，[33] では，オフロードに適切なループ文探索はできるが，ループ毎に CPU と GPU のデータ転送が発生する場合等効率的でない場合や，オフロードしても高速にならない場合もある等，高速化できるアプリケーションが限られていた．

そこで，更に，非効率なデータ転送を低減するため，明示的指示行を用いたデータ転送指定を，GA での並列処理の抽出と合わせて行う事を提案する．提案方式は，GA で生成された各個体について，ループ文の中で利用される変数データの参照関係を分析し，ループ毎に毎回データ転送するのでなくループ外でデータ転送して良いデータについては，ループ外でのデータ転送を明示的に指定する．

データ転送の種類は，CPU から GPU へのデータ転送，及び，GPU から CPU へのデータ転送がある．

・CPU から GPU へのデータ転送．

CPU プログラム側で定義した変数と GPU プログラム側で参照する変数が重なる場合は，CPU から GPU へのデータ転送が必要として，データ転送指定を行う．データ転送を指定する位置は，GPU 処理するループ文かそれより上位のループ文で，該当変数の設定，定義を含まない最上位のループとする．データ転送指示行の挿入位置は，for, do, while 等のループの直前に行う．

・GPU から CPU へのデータ転送．

GPU プログラム側で設定した変数と CPU プログラム側で参照する変数とが重なる場合は，GPU から CPU へのデータ転送が必要として，データ転送指定を行う．データ転送を指定する位置は，GPU 処理するループ文か，それより上位のループ文で，該当変数の参照，設定，定義を含まない最上位のループとする．データ転送指示行の挿入位置は，for, do, while 等のループの直前に行う．

このように，出来るだけ上位のループでデータ転送を一括して行うように，データ転送を明示的に指示することで，ループ毎に毎回データを転送する非効率な転送を避けることができる．

更に，gcov [35]，gprof [36] 等のプロファイリングツールを用いて，ループ回数を事前にチェックし，GPU オフロードを試行開始するかの振り分けをしても良い．

まとめると，GPU 向けオフロードについては，ループ文の GA 等を用いた適切なオフロード部の探索と，変数参照関係を用いたデータの一括転送により，自動での高速化を行える．

**4.1.2　FPGA 向けオフロード**

GPU では高速化は，ループ文等の並列処理が中心であった．一方，FPGA では，並列処理とパイプライン処理を活用して高速化するのが一般的であり，オフロードの自由度は高いが，機械がオフロード用のロジックを自動生成することは難しいのが現状である．そこで，FPGA のオフロードでは，今までにプログラマーが蓄積したノウハウ（Well-known パターン）を生か

— 4 —

して，大きな単位でオフロードする [37].

具体的には，Step1 のコード分析で把握した，類似コード検出等を用いたコード機能ブロックが FFT 処理である場合や，ライブラリ呼び出しで FFT 処理を呼び出している場合に，FFT 処理で既に定義されている FPGA ロジックに置換して，オフロードを行う．この置換のために，コードパターン DB は，FPGA にオフロード可能な処理のライブラリ呼び出しや機能ブロックと，オフロードする FPGA の処理ロジックを OpenCL や HDL で記述したコードを，登録している．環境適応機能は，コード分析結果と，コードパターン DB を照合し，オフロード可能な処理を，FPGA にオフロードする処理ロジック記述に置換する．

### 4.2 リソース量調整

Step4 のリソース量調整については，まず適切なリソース比を決め，次に性能，コスト要件に合うリソース量に設定する．アプリケーションを CPU と GPU で動作させるようオフロードするコードに Step1-3 で変換したとして，コード自体は適切であっても，CPU とオフロード先である GPU とのリソース量が適切なバランスでない場合は，性能が出ない．例えば，ある処理を行う際に，CPU の処理時間が 1000 秒，GPU の処理時間が 1 秒では，CPU がボトルネックとなるっている．[38] では，CPU と GPU を使って MapReduce フレームワークで処理している際に，CPU と GPU の実行時間が同じになるよう Map タスクを配分することで，全体の高性能化を図っている．私達も，リソース比を決める際は，[38] 等を参考に，何れかのハードウェアでの処理がボトルネックとなる配置を避けるため，想定するテストケースの処理時間から，CPU とオフロード先の処理時間が同等オーダーになるよう，リソース比を決定する．

リソース比を決定後は，想定するテストケースの処理性能が，Step1 でユーザが指定した要求性能及びコスト要求を満たすように，リソース比はキープして，リソース量を決定する．

### 4.3 配置場所調整

Step5 の配置場所調整では，性能，コストが適切になる場所を計算し配置先を決める．適切な配置先を決める手法については，最適化計算を用いた手法を検討している [?]．配置先を決めるための情報としては，配置するアプリケーションの想定するテストケースの性能情報（処理遅延やスループット）と，システムで利用できる設備リソース情報（クラウド，エッジ，Home GW 等の計算リソース，ノード間帯域，及びその既に利用されている量と，利用した際のコスト）がある．

配置先を決定するロジックは以下のようになる．想定テストケースの性能結果から，アプリケーションを配置した際の計算量と発生トラフィックを算出する．合わせて，クラウド，エッジ，Home GW 等のリンク関係をモデル化しておく．アプリケーションを特定のノードに配置した際に，コストが要求条件に収まることを制約条件に，処理遅延やスループット等の性能を最大化する配置，あるいは性能が要求条件を満たす形でコストが最低になる配置を，線形計画手法等を用いて導く．

### 4.4 動 作 検 証

Step6 の動作検証では，Step1-3 で決まった実行ファイルを Step4 の指定リソース量で Step5 の指定場所に配置した後に，期待通りの動作であることを，性能検証テストケースやリグレッションテストケースを実行することで確認する．性能検証テストケースは，ユーザが指定した想定テストケースを Jenkins [39]Selenium [40] 等の試験自動実行ツールを用いて行い，処理時間やスループット等を測定する．リグレッションテストについては，システムにインストールされるミドルウェアや OS 等のソフトウェアの情報を取得して，それらに対応するリグレッションテストを Jenkins 等を用いて実行する自動検証技術が検討されており [41]，それらを用いる．

### 4.5 運用中再構成

アプリケーション運用にて，Step6 の運用開始後，リクエスト特性の変化等で当初期待していた性能が出ない場合に，環境適応機能は，ソフトウェア設定，ソフトウェア/ハードウェア構成を再構成する．再構成の判断は，運用開始前に想定したテストケースでなく，現在の実運用にマッチするテストケースを元に，Step1-5 のコード変換，リソース量調整，配置場所調整を試行模擬し，性能，コストが，ユーザの期待を満たす場合に，再構成をユーザに提案し，了承後，再構成する．

#### 4.5.1 ソフトウェア設定

ソフトウェア設定の変更は，Step4-5 の処理を，周期的，または，性能がある閾値以下となった場合に試行模擬し，性能向上やコスト低減度合を計算する．リソース量の変更や配置場所の変更で性能やコストが改善できる見込みがある場合は，ユーザに再構成を提案する．ユーザ了承を得て，再構成を実施する場合に，リソースを変更する際は，クラウド関連技術により，メモリ量等の変更であれば断時間は無く変更できることが多い．

配置場所変更の際は，一括プロビジョニング技術（OpenStack [42] Heat を使った手法等 [43]）を用いて，移行先環境を複製しておき，そこに移行元からマイグレーションを行う．

#### 4.5.2 ソフトウェア/ハードウェア構成

ソフトウェア/ハードウェア構成の変更は，Step1-3 の処理を，周期的，または，性能がある閾値以下となった場合に試行模擬し，コード変換して，GPU オフロードのソフトロジック変更や FPGA のハードロジックの変更 [44] で，性能やコストが改善できる見込みがある場合は，ユーザに再構成を提案する．ユーザ了承を得て，再構成を実施する際に，GPU オフロードするソフトロジックの変更等，ソフトウェア構成の変更の場合は，更新する実行ファイルを起動する環境を複製後，アプリケーション実行中のデータをマイグレーションする．

再構成する際に，FPGA 等のハードロジックを変更する場合は，FPGA のハードロジックを運用中に再構成する．FPGA のハードロジック構成の再構成は，近年の Altera, Xilinx のツール [45] を用いると，運用中の数秒単位での再構成が可能である．

## 5. ま と め

本稿では，GPU, FPGA, IoT デバイス等環境が多様になる中で，アプリケーションを環境に合わせて適応させ，GPU や FPGA 等を適切に活用し，高性能にアプリケーションを動作させるための環境適応型ソフトウェアを提案した．環境適応型ソフトウェアにより，一度書いたソフトウェアを，異なる環境で

も高性能に動作させることを狙う．

環境適応型ソフトウェアは，アプリケーションの実行依頼に基づき，処理フローとして，コード変換，リソース量調整，配置場所調整，動作検証を行って，アプリケーションを実稼働させる．更に運用中に，ソフトウェア設定，ソフトウェア/ハードウェア構成の再構成を行う事で，リクエスト数変化等の状況変化に対応する．環境適応型ソフトウェアの個々の処理を実現する要素技術の基本検討を行った．

## 文　　献

[1] AWS EC2 web site, https://aws.amazon.com/ec2/instance-types/
[2] Y. Yamato, et al., "Fast and Reliable Restoration Method of Virtual Resources on OpenStack," IEEE Transactions on Cloud Computing, Sep. 2015.
[3] Y. Yamato, et al., "Software Maintenance Evaluation of Agile Software Development Method Based on OpenStack," IEICE Transactions on Information & Systems, Vol.E98-D, No.7, pp.1377-1380, July 2015.
[4] Y. Yamato, "OpenStack Hypervisor, Container and Baremetal Servers Performance Comparison," IEICE Communication Express, Vol.4, No.7, pp.228-232, July 2015.
[5] A. Putnam, et al., "A reconfigurable fabric for accelerating large-scale datacenter services," ISCA'14, pp.13-24, 2014.
[6] J. Sanders and E. Kandrot, "CUDA by example : an introduction to general-purpose GPU programming," Addison-Wesley, 2011
[7] J. E. Stone, et al., "OpenCL: A parallel programming standard for heterogeneous computing systems," Computing in science & engineering, Vol.12, No.3, pp.66-73, 2010.
[8] M. Hermann, et al., "Design Principles for Industrie 4.0 Scenarios," Rechnische Universitat Dortmund. 2015.
[9] Tron project web site, http://www.tron.org/
[10] Y. Yamato, et al., "Analyzing Machine Noise for Real Time Maintenance," 2016 8th International Conference on Graphic and Image Processing (ICGIP 2016), Oct. 2016.
[11] P. C. Evans and M. Annunziata, "Industrial Internet: Pushing the Boundaries of Minds and Machines," Technical report of General Electric (GE), Nov. 2012.
[12] Y. Yamato, "Proposal of Vital Data Analysis Platform using Wearable Sensor," 5th IIAE International Conference on Industrial Application Engineering 2017 (ICIAE2017), pp.138-143, Mar. 2017.
[13] Y. Yamato, "Ubiquitous Service Composition Technology for Ubiquitous Network Environments," IPSJ Journal, Vol.48, No.2, pp.562-577, Feb. 2007.
[14] Y. Yamato, et al., "Context-aware Ubiquitous Service Composition Technology," The IFIP International Conference on Research and Practical Issues of Enterprise Information Systems (CONFENIS 2006), pp.51-61, Apr. 2006.
[15] Y. Nakano, et al., "Web-Service-Based Avatar Service Modeling in the Next Generation Network," the 7th Asia-Pacific Symposium on Information and Telecommunication Technologies (APSITT2008), pp.52-57, Apr. 2008.
[16] M. Takemoto, et al., "Service Elements and Service Templates for Adaptive Service Composition in a Ubiquitous Computing Environment," APCC 2003, Vol.1, 2003.
[17] Y. Yamato, "Method of Service Template Generation on a Service Coordination Framework," 2nd International Symposium on Ubiquitous Computing Systems (UCS 2004), Nov. 2004.
[18] Y. Yamato, et al., "Study and Evaluation of Context-Aware Service Composition and Change-Over Using BPEL Engine and Semantic Web Techniques," IEEE Consumer Communications and Networking Conference (CCNC 2008), pp.863-867, Jan. 2008.
[19] Y. Yamato and H. Sunaga, "Abstract Service Scenario Generation Method for Ubiquitous Service Composition," IEICE Transactions on Communications, Vol.J91-B, 2008.
[20] Y. Yamato and H. Sunaga, "Context-Aware Service Composition and Component Change-over using Semantic Web Techniques," IEEE ICWS 2007, pp.687-694, July 2007.
[21] Y. Yokohata, et al., "Context-Aware Content-Provision Service for Shopping Malls Based on Ubiquitous Service-Oriented Network Framework and Authentication and Access Control Agent Framework," IEEE CCNC 2006, pp.1330-1331, 2006.
[22] J. Gosling, et al., "The Java language specification, third edition," Addison-Wesley, 2005. ISBN 0-321-24678-0.
[23] S. Wienke, et al., "OpenACC-first experiences with real-world applications," Euro-Par Parallel Processing, 2012.
[24] M. Wolfe, "Implementing the PGI accelerator model," ACM the 3rd Workshop on General-Purpose Computation on Graphics Processing Units, pp.43-50, Mar. 2010.
[25] K. Ishizaki, "Transparent GPU exploitation for Java," CANDAR 2016, Nov. 2016.
[26] E. Su, et al., "Compiler support of the workqueuing execution model for Intel SMP architectures," In Fourth European Workshop on OpenMP, Sep. 2002.
[27] H. Sunaga, et al., "Service Delivery Platform Architecture for the Next-Generation Network," ICIN2008, 2008.
[28] Y. Yamato, et al., "Development of Service Control Server for Web-Telecom Coordination Service," IEEE ICWS 2008, pp.600-607, Sep. 2008.
[29] Y. Yokohata, et al., "Service Composition Architecture for Programmability and Flexibility in Ubiquitous Communication Networks," IEEE International Symposium on Applications and the Internet Workshops (SAINTW'06), 2006.
[30] Clang website, http://llvm.org/
[31] CCFinder web site, http://www.ccfinder.net/
[32] OpenCV web site, http://opencv.org/
[33] Y. Yamato, et al., "Automatic GPU Offloading Technology for Open IoT Environment," IEEE Internet of Things Journal, Sep. 2018.
[34] J. H. Holland, "Genetic algorithms," Scientific american, Vol.267, No.1, pp.66-73, 1992.
[35] gcov website, http://gcc.gnu.org/onlinedocs/gcc/Gcov.html
[36] gprof website, http://sourceware.org/binutils/docs-2.20/gprof/
[37] Y. Yamato, "Optimum Application Deployment Technology for Heterogeneous IaaS Cloud," Journal of Information Processing, Vol.25, No.1, pp.56-58, Jan. 2017.
[38] K. Shirahata, et al., "Hybrid Map Task Scheduling for GPU-Based Heterogeneous Clusters,"IEEE CloudCom, pp.733-740, Dec. 2010.
[39] Jenkins web site, https://jenkins.io/
[40] Selenium web site, https://www.seleniumhq.org/
[41] Y. Yamato, "Automatic verification technology of software patches for user virtual environments on IaaS cloud," Journal of Cloud Computing, Springer, 2015, 4:4, Feb. 2015.
[42] O. Sefraoui, et al., "OpenStack: toward an open-source solution for cloud computing," International Journal of Computer Applications, Vol.55, 2012.
[43] Y. Yamato, "Performance-Aware Server Architecture Recommendation and Automatic Performance Verification Technology on IaaS Cloud," Service Oriented Computing and Applications, Springer, Nov. 2016.
[44] Y. Yamato, "Server Selection, Configuration and Reconfiguration Technology for IaaS Cloud with Multiple Server Types," Journal of Network and Systems Management, Springer, Aug. 2017.
[45] Altera SDK web site, https://www.altera.com/products/design-software/embedded-software-developers/opencl/documentation.html